\documentclass[aps,pre,reprint,groupedaddress,twocolumn]{revtex4-2}

\usepackage{graphicx}
\usepackage{dcolumn}
\usepackage{bm}
\usepackage{amsmath,amssymb,subfigure}
\usepackage{color}
\usepackage[normalem]{ulem}
\usepackage{epstopdf}
\usepackage{esint}

\newcommand{\reffig}[1]{Fig.~\ref{#1}}
\newcommand{\eq}[1]{Eq.~\eqref{#1}}

\newcommand{\vtheta}{\bar{v}_\theta^\ast} 
\newcommand{\vz}{\bar{v}_z^\ast} 
\newcommand{\omegatheta}{\bar{\Omega}_\theta^\ast} 
\newcommand{\omegaz}{\bar{\Omega}_z^\ast} 
\newcommand{\Hel}{\mathcal{H}^\ast} 
\newcommand{\hel}{\bar{\mathfrak{h}}^\ast} 

\begin{document}
\title{Contactless generation and trapping of hydrodynamic knots in sessile droplets by acoustic screw dislocations}
\author{Shuren Song, Jia Zhou, Antoine Riaud}
\email{antoine\_riaud@fudan.edu.cn}
\affiliation{\\State Key Laboratory of ASIC and System, \\School of Microelectronics, \\Fudan University, \\Shanghai, P. R. China}

\author{Antonino Marcian\`o}
\affiliation{Department of Physics \& Center for Field Theory and Particle Physics, Fudan University, 200433 Shanghai, China}
\affiliation{Laboratori Nazionali di Frascati INFN Via Enrico Fermi 54, Frascati (Roma), Italy, EU}

\date{\today}
\begin{abstract}
	Hydrodynamics knots are not only promising elementary structures to study mass and momentum transfer in turbulent flows, but also potent analogs for other topological problems arising in particle physics. However, experimental studies of knots are highly challenging due to the limited control over knot generation and difficult observation of the resulting fast-paced multiscale flow evolution. In this paper, we propose using acoustic streaming to tie hydrodynamic knots in fluids. The method is contactless, almost instantaneous and is relatively insensitive to viscosity. Importantly, it allows starting from quite arbitrary three dimensional flow structures without relying on external boundary conditions. We demonstrate our approach by using an acoustic screw dislocation to  tie a knot in a sessile droplet. We observe an inversion of the knot chirality (measured by the hydrodynamic helicity) as the topological charge of the screw dislocation is increased. Combined with recent progress in acoustic field synthesis, this work opens a window to study more complex hydrodynamic knot topologies at a broader range of space and timescales.
\end{abstract}
\maketitle
\section{Introduction}
\begin{figure*}
	\includegraphics[width=5in]{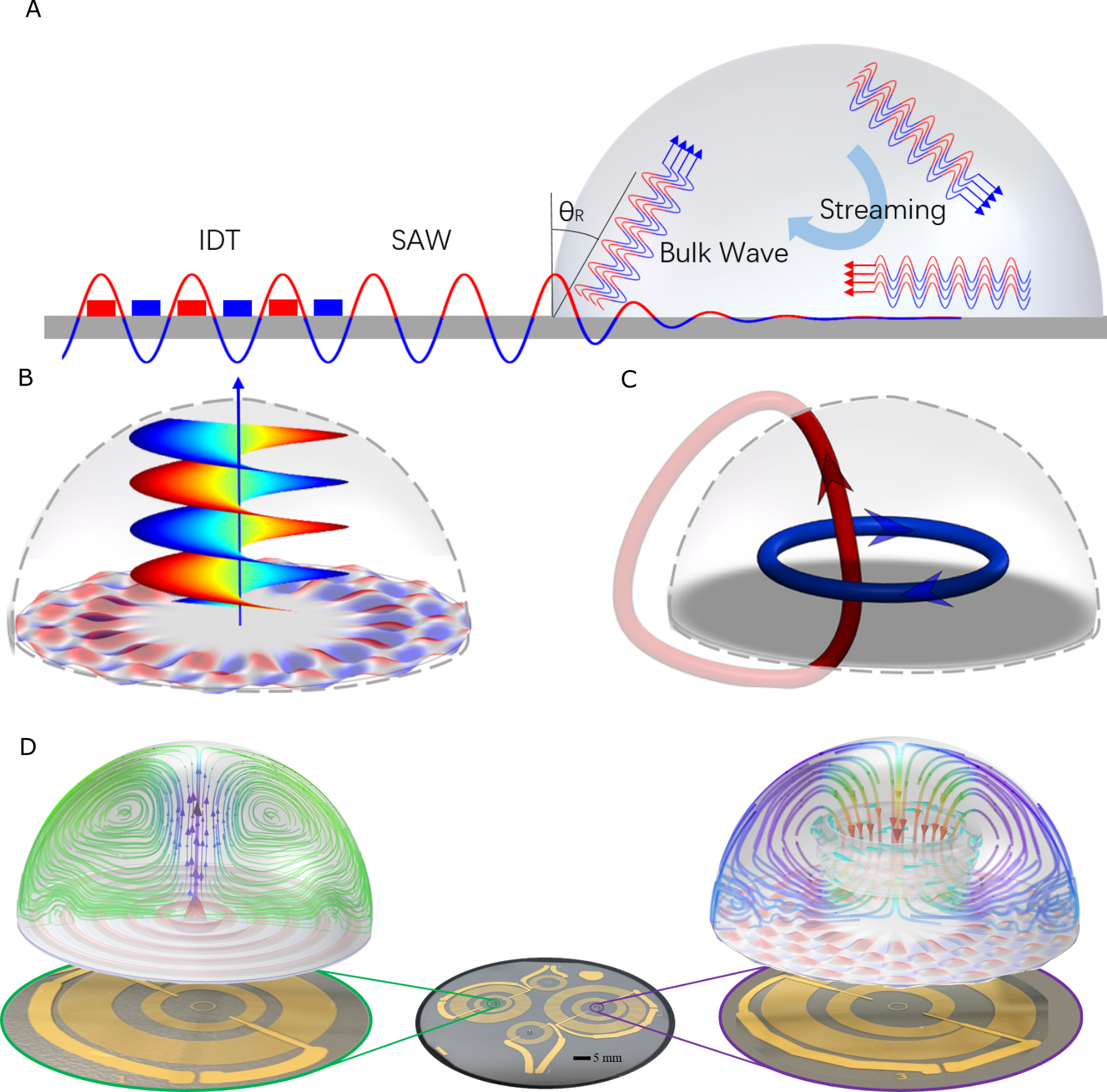}	
	\caption{Experiment principle. \textbf{(A)} Generation of acoustic streaming: an interdigitated transducer on a piezoelectric surface launches as surface acoustic wave. The wave travels along the substrate surface until it meets a liquid in which it leaks as a bulk acoustic wave. The wave then transfers its momentum to the fluid which generates a steady acoustic streaming flow. \textbf{(B)} isophase of an acoustic vortex traveling in a sessile droplet. The incident swirling SAW is shown as an undulation at the bottom of the droplet. \textbf{(C)} Hydrodynamic knot produced by two hydrodynamic vortex filaments. \textbf{(D)} Schematic of the experimental setup: two sets of spiraling IDTs, shown in the center, generate swirling SAWs $\mathcal{W}_0$ ($\mathcal{W}_{15}$) on the left (right) that imprint a poloidal (helical) flow motion.}
	\label{fig: setup}
\end{figure*}
Knots encode 3-dimensional physical fields in the topology of simple constructs such as screw dislocations \cite{berry2001knotted} or vortex filaments \cite{kleckner2013creation}. In fluid mechanics, knots are measured in terms of helicity which, similarly to energy or momentum, remains invariant in isobaric inviscid fluids \cite{moreau1961constantes,moffatt1969degree}. Knots and their helicity are regarded as promising elementary constituents of turbulence \cite{moffatt2014helicity} and could play a key role in turbulent mass and momentum transfer, but also in particle and bubbles aggregation \cite{squires1991preferential,maxey1987gravitational,douady1991direct}. Yet, observing vortex knots is a daunting task \cite{wallace1992experimental,eidelman2014helicity,stepanov2018analysis} due to (i) the difficulty to reliably generate single hydrodynamic knots and (ii) the need to quickly measure the three-dimensional flow that surrounds these knots in order to obtain the helicity. 

This generation challenge was recently addressed by imprinting a helical motion to water using specially designed hydrofoils \cite{kleckner2013creation,scheeler2017complete}. This method comes with several challenges: first, the vortex escapes and drifts with the fluid, which requires a large experimental domain to let the vortex evolve. This in turn limits the spatial resolution of the flow, such that the experimenter must assume that the vortex filaments are infinitely thin to estimate the helicity. Second, in real fluids, helicity is not conserved which prevents steady-state experiments. Third, the hydrofoils must contact the fluid at the beginning of the experiment, which may become an issue when the choice of material has an influence on the experiment, such as in geodynamo models \cite{moffatt2014helicity}. Fourth, the generated vortex knot is necessarily large in order to reach a high Reynolds number ($\simeq20,000$) \cite{scheeler2017complete} to ensure a sufficient lifetime for the knots. This precludes investigating and using vortex knots in microfluidic systems.
 
Besides hydrofoils and other solid walls, vorticity can be generated by acoustic streaming. Herein, the momentum flux of an acoustic wave is transferred to the fluid as the wave attenuates in the bulk (Eckart) \cite{eckart1948vortices} or by contact with the walls (Rayleigh) \cite{rayleigh1884circulation}, thereby imprinting a steady flow motion to the fluid. Compared to the hydrofoils, acoustic streaming is contactless, relatively insensitive to fluid viscosity and its force can be turned on and off instantly compared to hydrodynamic timescales. While coupling of weak capillary or acoustic waves with intense hydrodynamic vortex filaments has inspired a wealth of elegant experiments in topological physics \cite{berry1980wavefront,roux1997aharonov,yang2015topological,aubry2019topological}, the reverse approach of finite-amplitude waves generating a flow has received less attention \cite{riaud2014cyclones,hong2015observation}. 

It was theoretically shown that acoustic screw dislocations (acoustic vortices) \cite{nye1974dislocations,durnin1987diffraction} propagating in infinitely long tubes can generate an acoustic streaming flow reminiscent of a cyclone  \cite{riaud2014cyclones}. Such cyclone is an elementary knot between an axial hydrodynamic vortex filament and an azimuthal hydrodynamic vortex ring. The same theory also predicted that the topological order $\ell$ of the screw dislocation (that is, the number of interfering wavefronts at the dislocation) dictates the axial direction of the flow. However, the three-dimensional structure of the hydrodynamic knot and the connections between this knot and the acoustic vortex topological order have never been verified experimentally.

In this paper, we experimentally demonstrate the synthesis of a microscale hydrodynamic knot in a sessile droplet using the acoustic streaming generated by an acoustic vortex. The generation is fast, contactless, highly reproducible and the knots can be sustained as long as the experimenter sees fit. We then measure the steady-state flow-field in three-dimensions by general defocusing particle tracking (GDPT) \cite{barnkob2015general,barnkob2020general}. Our measurements agree with numerical simulations and show that helicity is confined within the acoustic vortex core and can be reversed for high enough topological order, thereby illustrating the effect of walls on flow topology.  
 
\section{Results}

\subsection{Experiment principle}

The cyclone-like flow in the droplet is driven by the acoustic streaming of an acoustic vortex. The helical wavefront of such screw dislocation is shown in \reffig{fig: setup}B. We consider a fluid of density at rest $\rho_0$, kinematic viscosity $\nu$, kinematic bulk viscosity $\nu'$ and sound velocity $c_0$ subject to an acoustic field of pulsation $\omega$, wavenumber $k_0$, pressure $\tilde{p}$ and vibration velocity $\mathbf{\tilde{v}}$. Analytical expressions of the acoustic streaming field of steady velocity $\mathbf{\bar{v}}$ and steady vorticity $\mathbf{\bar{\Omega}}$ can be derived by perturbation expansion when the acoustic pressure is not too large ($\tilde{p}\ll\rho_0{c_0}^2$) and the acoustic attenuation is moderate  ($\frac{{c_0}^2}{\omega\nu b} \gg 1$),  where $b=\frac{4}{3}\nu'/\nu$. This allows successively solving fluid quantities of a field $x$ first at rest $x_0$, then as small acoustic quantities $\tilde{x}$ and finally very small steady state effects of nonlinear acoustics $\bar{x}$. Upon such perturbation, acoustic streaming can be modeled as a steady solenoidal acoustic streaming force $\mathbf{F}$ driving a Stokes flow \cite{eckart1948vortices,riaud2017influence}:
\begin{subequations}
	\begin{align}
	\nu \nabla \times \mathbf{\bar{\Omega}} &= \mathbf{F},   \label{eq: stokes_stream}\\
	\text{with: } &\mathbf{F} = \frac{\omega^2\nu b \langle\mathbf{\Pi}\rangle}{c_0^4},    \label{eq: Feckart}
	\end{align}
\end{subequations}
where $\langle  \mathbf{\Pi} \rangle$ indicates the time-average of the acoustic power flux $\mathbf{\Pi} = \tilde{p}\mathbf{\tilde{v}}$. We note that while acoustic streaming can develop into high-Reynolds number turbulent jets \cite{lighthill1978acoustic}, the theory is rigorously tractable only for creeping flows as above. Quite interestingly, the acoustic streaming force is set only by the acoustic field, which can be precisely controlled in three dimensions by holographic techniques \cite{riaud2015anisotropic,marzo2015holographic,melde2016holograms}. By using the acoustic streaming to generate a steady flow, and then turning the acoustic field off, one can follow the spontaneous (non-forced) evolution of an arbitrary flow structure. Reversely, one can keep the acoustic forcing on at all times to generate a steady flow as it will be done in the remaining of the paper.

When the acoustic wave is a vortex, the acoustic streaming force combines a pushing force and a torque \cite{anhauser2012acoustic} proportional to the topological order $\ell$ \cite{thomas2003pseudo}, which generates a cyclone-like flow with the knotted pair of vortex filaments shown in \reffig{fig: setup}C.

\subsection{Generation of the acoustic vortex}

\begin{figure*}
	\includegraphics[width=6in]{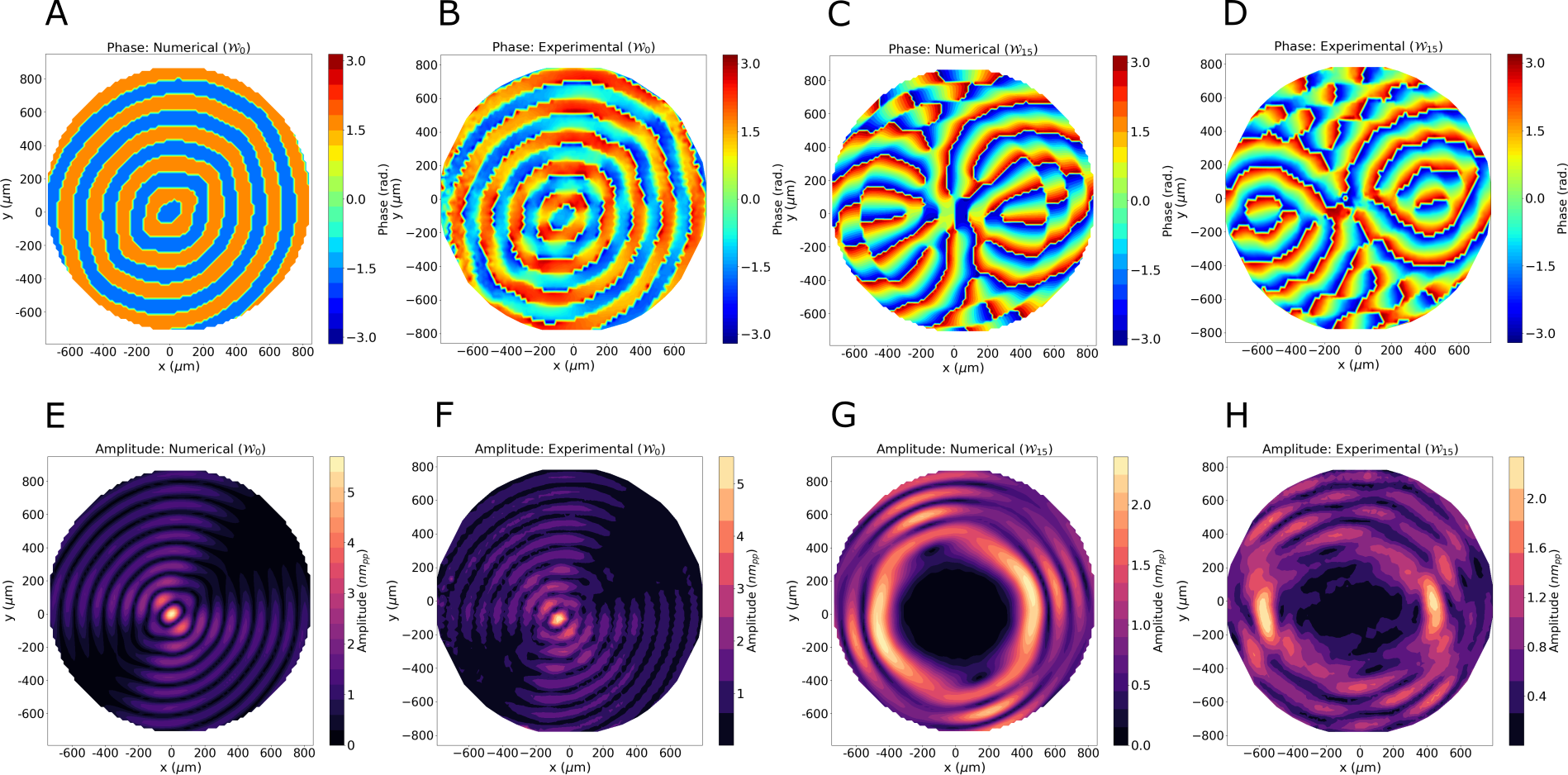}
	\caption{Vibration phase and amplitude of the incident swirling SAW $\mathcal{W}_0$ \textbf{(left)} and $\mathcal{W}_{15}$ \textbf{(right)}. \textbf{(A, C, E, G)}: theoretical fields computed from \eq{eq: aniso_vortex}. \textbf{(B, C, D, H)}: experimental vibration fields obtained by laser vibrometry. The maximum amplitude is 5.5 nm in for the $\mathcal{W}_0$ \textbf{(F)} and 2.3 nm for the $\mathcal{W}_{15}$ \textbf{(H)}.}
	\label{fig: scans}
\end{figure*}

Acoustic vortices can be generated by diffraction networks , metamaterials \cite{jiang2016convert,naify2016generation}, transducer arrays \cite{hefner1999acoustical,thomas2003pseudo,riaud2015anisotropic} and spiraling interdigitated transducers (IDT) patterned on a piezoelectric substrate \cite{riaud2017selective,baudoin2019folding}. Here, we choose the latter approach for its straightforward miniaturization, easy scale-up and simpler integration in microfluidic systems, which may enable well-controlled experiments on vortex knots and their networks at microscale. The acoustic streaming generation principle is sketched in \reffig{fig: setup}A and the setup is shown in \reffig{fig: setup}D. Briefly, a spiraling interdigitated transducer patterned on a piezoelectric substrate generates a swirling surface acoustic wave (SAW). When a droplet is placed on the transducer, this surface wave radiates into the liquid to become a bulk acoustic wave dislocation, which then drives the acoustic streaming. Accounting for the piezoelectric substrate anisotropy, the vertical displacement of the swirling SAW reads:
\begin{equation}
	\mathcal{W}_\ell(\mathbf{r}) = \frac{i^\ell e^{i\omega t}}{2\pi }\int_{-\pi}^{+\pi} a(\psi)e^{-i\ell\psi-i\mathbf{k}(\psi)\cdot\mathbf{r}}d\psi,\label{eq: aniso_vortex}
\end{equation}
with $\mathbf{k}(\psi)$ and $a(\psi)$ the wavevector and complex amplitude of a plane surface acoustic wave generated by a straight IDT \cite{laude2008subwavelength,riaud2017selective}, and $\mathbf{r}$ the position vector relatively to the vortex center. In the isotropic case, \eq{eq: aniso_vortex} simplifies into $W_\ell = aJ_\ell(k_r r)e^{i\omega t-i\ell\theta}$, with $k_r$ the wavenumber of the SAW. $\ell$ is the acoustic vortex topological order, that is the number of intertwined acoustic wavefronts on the vortex axis. These wavefronts destructively interfere to create a hollow core surrounded by a bright ring with a radius $r_\text{hole}\simeq \frac{1}{k_r}[\ell+0.809\ell^{1/3}]$. By analogy with the infinite tube case, we infer that the axial flow reversal can only be observed when the hole radius is similar to that of the droplet. Given the 20 MHz acoustic frequency chosen for efficient ultrasonic power transfer from the surface wave to the droplet \cite{connacher2018micro}, we fabricate two transducers with $\ell = 0$ (to demonstrate repeller vortices flowing away from the transducer) and $\ell = 15$ (to demonstrate attractor vortices flowing toward the transducer). 

The experimental phase and amplitude of the generated swirling SAW measured by laser vibrometry are shown in \reffig{fig: scans} side by side with their theoretical values from \eq{eq: aniso_vortex}. The $\mathcal{W}_0$ displays a sharply focused peak in good agreement with its theoretical expression, while the $\mathcal{W}_{15}$ is slightly less faithful in phase and amplitude, thereby attesting of the delicate interference pattern controlling the general vortex structure. Green function simulations \cite{laude2008subwavelength} of our transducers yield similar imperfect patterns (see SI), suggesting that the analytical equation in \cite{riaud2017selective} may not be suitable for such very high $\ell$ vortices, and that other optimization approaches \cite{tian2020generating} could fare better. Nonetheless, the $\mathcal{W}_{15}$ still display the wide acoustic hole deemed to be critical for flow reversal.
 
\subsection{Orders of magnitude}

We estimate the helicity density $\bar{\mathfrak{h}} = \bar{\mathbf{v}}\cdot\bar{\mathbf{\Omega}}$ based on the calculations for an unbounded tube \cite{riaud2014cyclones}. In this case, the axial and azimuthal flow velocity scale as $\vz\simeq \frac{\omegatheta}{k_r}$ and $\vtheta\simeq \frac{\omegaz}{k_r}$, respectively, where $\omegatheta$ and $\omegaz$ stand for the characteristic azimuthal and axial vorticity respectively. It is then remarkable that the contribution of $\vtheta\omegatheta$ and $\vz\omegaz$ to the helicity share exactly the same weight, which yields $\hel\simeq 2\frac{\omegatheta\omegaz}{k_r}$. Accounting for the pseudo-orbital angular momentum of the acoustic vortex, the axial vorticity reads $\omegaz \simeq \frac{\ell \omegatheta}{\tan\alpha}$ with $\tan\alpha = \frac{k_r}{\sqrt{{k_0}^2-{k_r}^2}}$ and $\omegatheta \simeq \frac{1}{2}\frac{\omega^3 b}{{c_0}^2}\tan\alpha \langle \tilde{u}^2\rangle$, where $\tilde{u}$ represents the characteristic vibration amplitude of the substrate. Therefore, the helicity should scale as $\Hel \simeq V_d\hel$ with:
\begin{equation}
	\bar{\mathfrak{h}}^\ast \simeq \frac{\ell b^2 \omega^5}{2{c_0}^3}\frac{\sin\alpha}{\cos^2\alpha}\langle \tilde{u}^2\rangle^2.
	\label{eq: Hest} 
\end{equation}
According to \eq{eq: Hest}, and similarly to acoustic streaming velocity, the flow helicity is relatively insensitive to the viscosity (but may depend on the acoustic attenuation \cite{riaud2017influence} as attested by the $b^2$ factor), and is very nonlinear with the acoustic vibration amplitude $\tilde{u}$. We also infer that the helicity is proportional to the topological order $\ell$. One may worry that the denominator $\cos\alpha$ may approach 0. This can only happen if the incident wave becomes a plane wave which would require $\ell = 0$ exactly (because $\ell$ must be an integer), thereby setting $\Hel=0$. The opposite case of $\sin\alpha=0$ is physically sound and represents an in-plane acoustic vortex that would not radiate in the fluid and thus would not yield any helicity.

\subsection{General defocusing particle tracking (GDPT)}
The experiment is similar to \cite{riaud2017influence} but uses a spiraling transducer instead of a straight one. Brielfly, the transducer is installed on the stage of inverted fluorescence microscope CKX53 (Olympus). To improve the GDPT performance, a cylindrical lens ($f = 1000$ mm) is placed under the stage. We then carefully drop a 2 $\mu$L droplet of 90 $w\%$  glycerin/water mixture containing 5 $\mu$m polystyrene particles for flow visualization. The mixture composition  is chosen to minimize the effects of acoustic radiation pressure (i) by matching the acoustic impedance of the particles \cite{augustsson2016iso} and (ii) by increasing the drag to radiation force ratio \cite{riaud2020mechanical,williams2017acoustophoretic}.

The GDPT is a tracking algorithm using spherical aberrations (such as those due to a cyclindrical lens) in order to estimate the depth position of the particles in addition to their in-plane location \cite{barnkob2015general,barnkob2020general,rossi2020fast}. In this study, we used the GDTPlab implementation \cite{barnkob2015general} but preliminary test indicate an improved performance with the newer DefocusTracker \cite{barnkob2021defocustracker}. After locating the particles, the velocity fields are reconstructed by numerical differentiation upon tracing a large number of particles. Videos of the particles motion are provided in SI.
 
\section{Discussion}

\subsection{Acoustic streaming velocity field}
\begin{figure*}
	\includegraphics[width=6in]{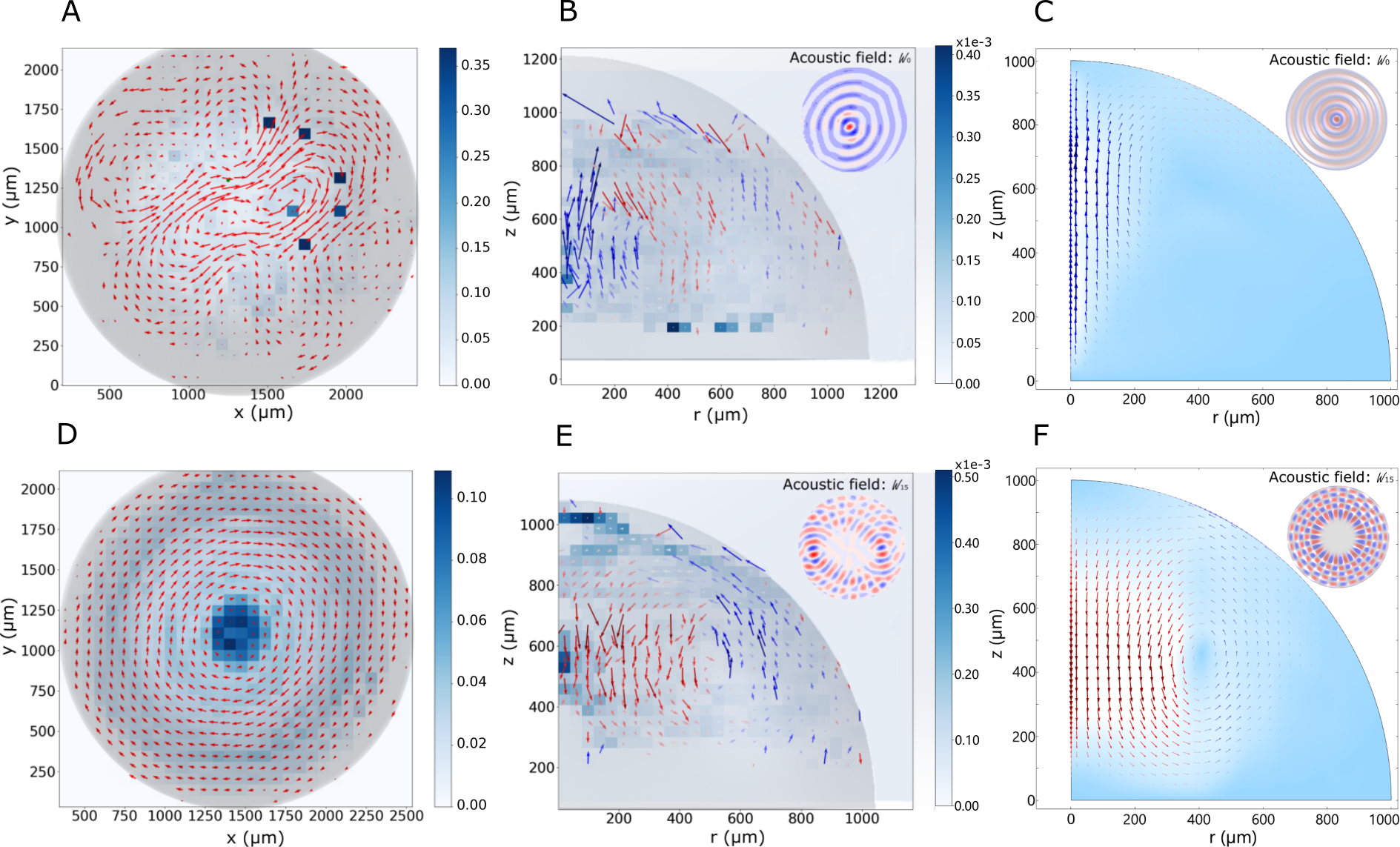}	
	\caption{Acoustic streaming flow in the droplet. \textbf{(Top)}: $\mathcal{W}_0$,  \textbf{(Bottom)}: $\mathcal{W}_{15}$. Azimuthal \textbf{(A,B)} and poloidal \textbf{(C,D)} components of the experimental hydrodynamic flows measured by GDPT. \textbf{(E,F)}: simulated acoustic streaming flow field. The arrows length is the velocity magnitude. The maximum in-plane velocities are 61 $\mu$m/s \textbf{(A)}, 46 $\mu$m/s \textbf{(B)}, 38 $\mu$m/s \textbf{(C)}, 588 $\mu$m/s \textbf{(D)}, 60 $\mu$m/s \textbf{(E)} and 51 $\mu$m/s \textbf{(F)}. The arrow colors in \textbf{(B,E)} indicate the direction (up/down) of the flow. The insets in \textbf{(B, C, E, F)}, show the incident swirling SAW in experiments (anisotropic) and simulations (isotropic). The background color in experimental figures \textbf{(A, B, D, E)} represents the number of observed particles in 1/$\mu$m$^2$ and 1/$\mu$m$^3$ over the recordings.}
	\label{fig: gdpt}
\end{figure*}
Bottom views of the droplet are shown in \reffig{fig: gdpt}A,D. The $\mathcal{W}_0$ generates no coherent azimuthal flow, which was expected due to symmetry constraints. Similarly to Hong et al. \cite{hong2015observation}, the  $\mathcal{W}_{15}$ generates a strong azimuthal flow driven by the torque of the acoustic vortex. We also note that more particles were observed at the droplet center, which indicates a longer residence time due to the stagnation point in this region \cite{raghavan2010particle}.

Using the GDPT image processing, we can reconstruct the three-dimensional flow in the droplet (see animation in SI) and visualize its cross-section (\reffig{fig: gdpt}B, E). We note that this direct observation of the flow field contrasts with earlier reports where the velocity field was reconstructed from the vortex filament structure using Biot and Savart circulation law \cite{kleckner2013creation}. The flow being essentially axisymmetric, the cross-sections are obtained by averaging the radial and axial velocities around the droplet. According to these reconstructed flow profiles, the $\mathcal{W}_0$ generates a poloidal upward flow (repeller vortex) while the  $\mathcal{W}_{15}$ drives a spiraling downward flow (attractor vortex), in agreement with the theory developed for unbounded tubes \cite{riaud2014cyclones}. 
 
These experimental results are compared to first-principle simulations (\reffig{fig: gdpt}C,F) obtained by solving \eq{eq: stokes_stream} and using the isotropic version of \eq{eq: aniso_vortex} as boundary condition, which allows using fast axisymmetric models instead of time-consuming three-dimensional ones \cite{riaud2017influence}. The good quantitative agreement between those simulations and the experiments using anisotropic wave fields suggests that the flow pattern is quite insensitive to the detailed excitation field. Having validated our simulations, we then use them (i) to visualize and calculate the flow helicity in the droplet, and (ii) to rapidly test the acoustic streaming generated by acoustic vortices over a range of topological orders, which would have been time-consuming experimentally.

\subsection{Distribution of helicity density}

\begin{figure}
	\includegraphics[width=2.8in]{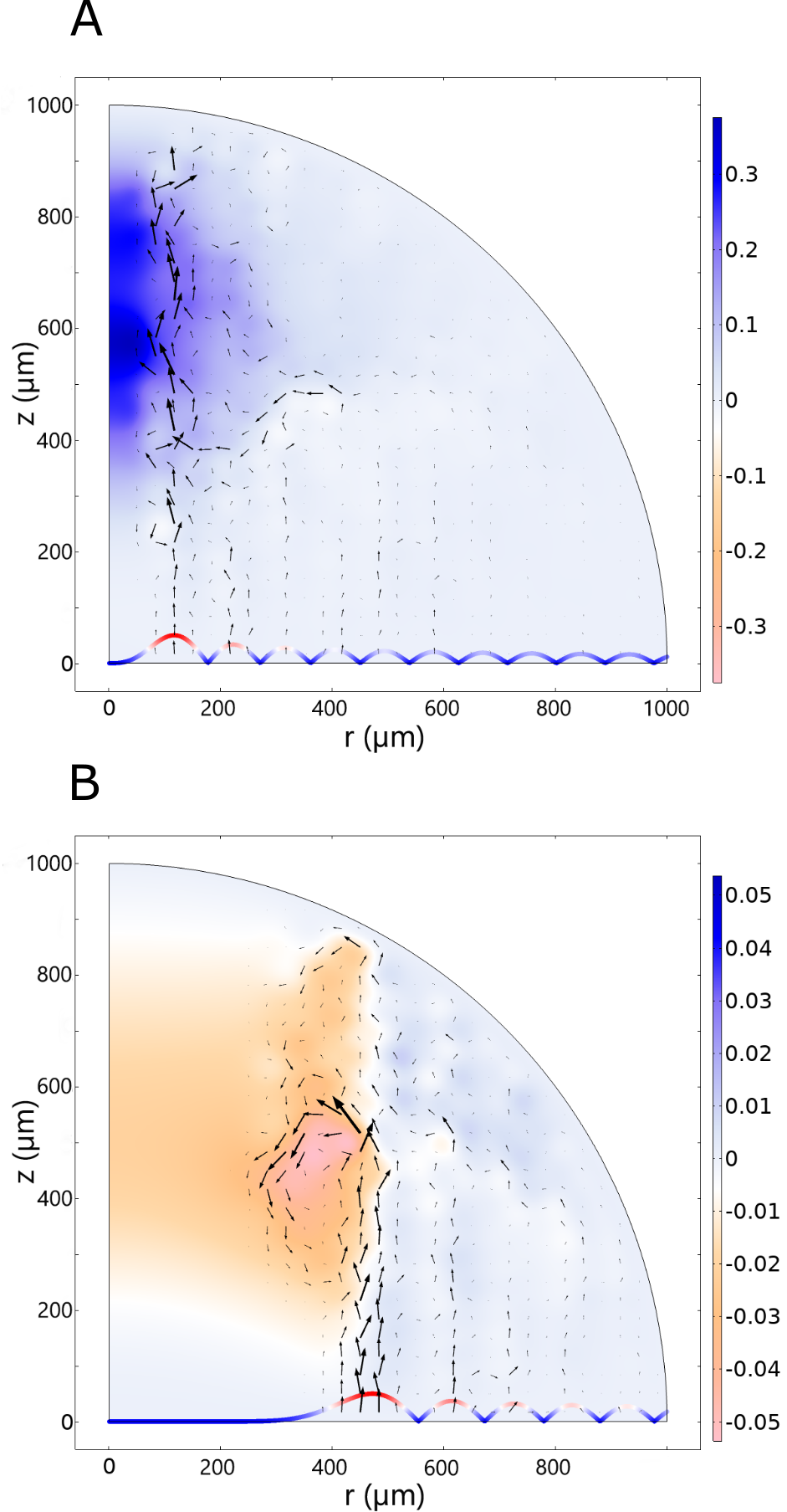}
	\centering
	\caption{Simulated helicity density ($\bar{\mathfrak{h}}/\hel$) for $\ell = 3$ \textbf{(A)} and $\ell = 15$ \textbf{(B)} in sessile droplets. The black arrows show the acoustic power flux density $\langle \mathbf{\Pi} \rangle$, and the colored curves at the bottom of the droplet illustrate the incident swirling SAW amplitude (in the isotropic case).}
	\label{fig: comsol_helicity}
\end{figure}

We first use our numerical model to compare the distribution of normalized helicity density $\bar{\mathfrak{h}}/\hel$ to the acoustic streaming force from \eq{eq: Feckart} in \reffig{fig: comsol_helicity}A, B for $\ell=3$ and $\ell=15$ respectively. In both cases, the helicity is confined in some blob located around the droplet axis where the streaming force is minimum (absence of black arrows). Recalling that the streaming force is proportional to the acoustic power flux, the force-free region corresponds to the hole of the acoustic vortex, suggesting that the extent of the helicity blob is confined to the vortex core. Interestingly, these simulations also show that the helicity changes sign when the topological charge increases. Physically, the force is mainly directed upwards but the closed geometry of the droplet forces the flow to recirculate either inside the vortex core or outside of it. At low $\ell$, the acoustic streaming force is concentrated in a small region around the central axis which forces the flow to recirculate outwards, whereas at high $\ell$ the force is stronger on the droplet periphery which drives the downward flow on the droplet axis. 

\subsection{Inversion of a hydrodynamic knot by increasing the acoustic screw dislocation topological order}   
 
 \begin{figure}
 	\includegraphics[width=2.8in]{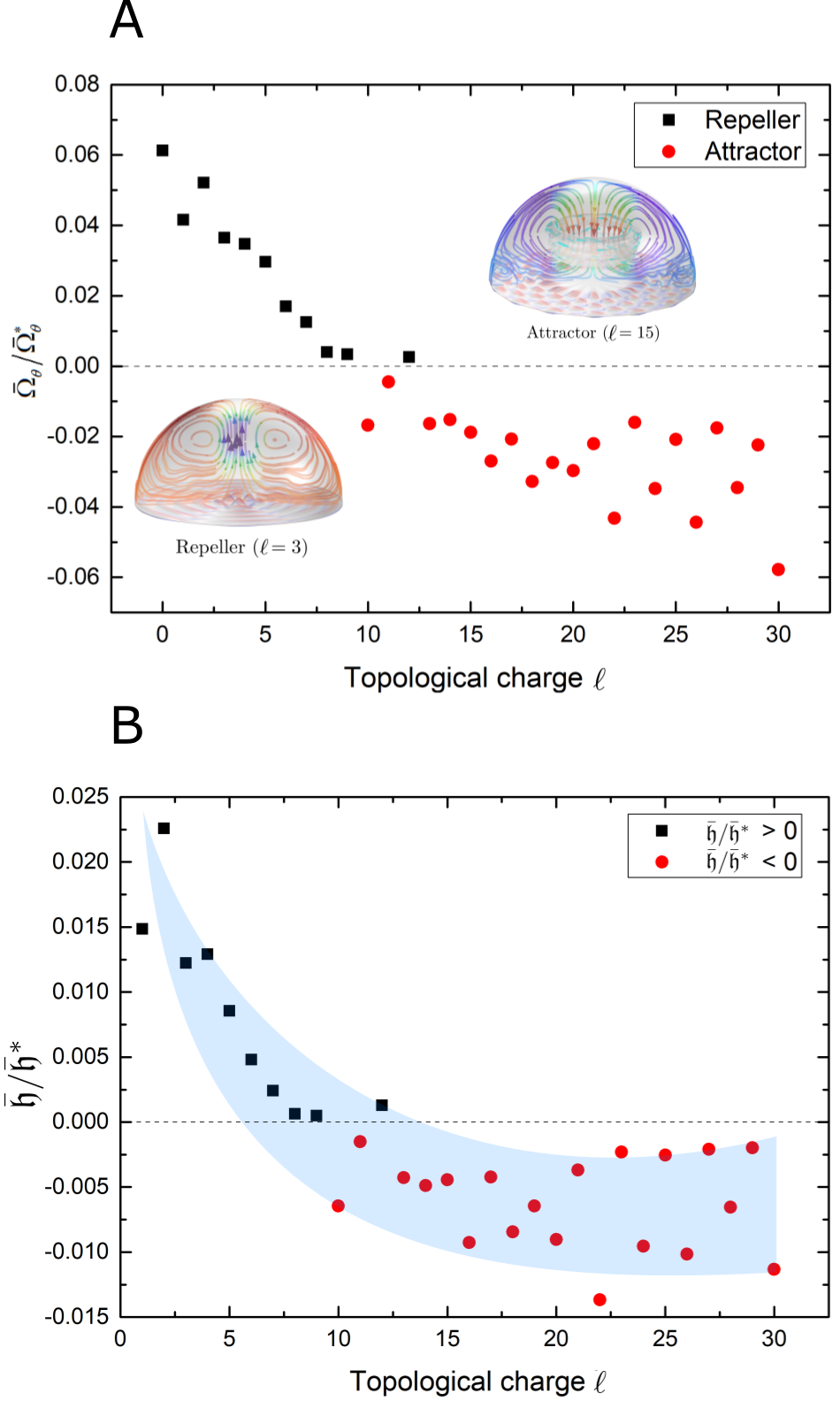}
 	\centering
 	\caption{Reversal of azimuthal vorticity $\bar{\Omega}_{\theta}$ and helicity $\bar{\mathfrak{h}}$ for increasing topological charge $\ell$.}
 	\label{fig: helicity_topocharge}
 \end{figure}

We then systematically investigate the sign change of the total hydrodynamic helicity by simulating $\bar{\mathcal{H}}$ for $\ell\in \left[0,30\right]$. The upper bound is chosen such that the vortex hole encompasses the whole droplet. The helicity (\reffig{fig: helicity_topocharge}B) follows the same trend as the average droplet vorticity $\frac{1}{V}\int_V\bar{\Omega}_\theta dV$ (shown in \reffig{fig: helicity_topocharge}A) with a flow reversal when $\ell\simeq 10$ marking the limit between attractor and repeller vortices. This illustrates that even interacting fields that accept an elementary topological description (here an acoustic screw dislocation and a hydrodynamic knot) can be coupled an in nontrivial way by the mean of boundary conditions. 

\subsection{Perspectives: connection to topological quantum field theory} 

Topological quantum field theory is a theoretical framework in physics that accounts for non-propagative collective degrees of freedom. The mathematical framework these theories are cast into is actually so wide that also non topological quantum field theories can be expressed as a ``deformation'' of topological quantum field theories, through the imposition of {\it ad hoc} constraints.

Furthermore, the states of the Hilbert space of topological theories are supported on lattices that are nothing but graphs, namely one-complexes provided with nodes and intertwined links. Depending on the local gauge group defining the topological theory, and on the associated connection form, one can introduce group elements called holonomies, which are the path ordered exponentials $\mathcal{P}$ of the integrated fluxes of the connection along the links, and invariant tensors intertwining holonomies. The former ones are assigned to the links of a graph, while the latter ones to the nodes. This prescription defines the so called spin-network basis.  

Denoting a link with $\gamma$, in our case holonomies $H_\gamma$ shall be calculated along the flux of the velocity field $\mathbf{v}$, which is an abelian connection, namely 
\begin{equation}
	H_{\gamma} (\mathbf{v})= \mathcal{P} \, e^{\int_\gamma \mathbf{v}} = e^{\int_\gamma \mathbf{v}}\,.
\end{equation}
Consequently, intertwiners associated to the nodes of the one-complexes are trivial $\mathbb{C}$-numbers. It is possible to find a map from the the spin-network basis to the loop basis \cite{rovelli1995spin}, and thus recast the theory in term of circuitation of the velocity field along loops. Since in our scenario loops are winding vortices, we may consider to associate flux lines to vortices, as orthogonal to the velocity flux lines (holonomies). Vortices may be then interpreted as the variables conjugated to velocities, and hence seen in terms of a loop-string duality \cite{addazi2020self}.

What makes topological quantum field theory extremely appealing, in particular for the purposes of the research direction we exposed here, is the connection to graph theory. Several relevant topological invariants, including the Jones polynomials \cite{witten1989quantum}, have been defined within topological quantum field theory in terms of their invariance properties under the Reidemeister moves, namely twist and untwist of loops/links, motion of loops/links one over another and motion of loops/links over and under crossings.

A very central role is played by the Chern-Simons theories. It was shown \cite{witten19882+} that the path-integral quantization of the Chern-Simons theory can be attained using non-perturbative methods. This partition function defines a topological invariant of a three-dimensional manifold, called Chern-Simons invariant, related to the Jones polynomials. A notable topological theory is three-dimensional General Relativity. For its version with cosmological constant there exist also quantization methods involving purely combinatorial and algebraic structures, while constructing the topological invariants.

A rigorous definition of topological invariant corresponding to the Chern-Simons classical theory was proposed by Reshetikhin and Turaev in \cite{reshetikhin1991invariants}, while the counterpart for the three-dimensional theory of gravity with cosmological constant was derived by Turaev and Viro \cite{turaev1992state}.
The mathematical framework we intend to implement in our planned analyses is precisely the aforementioned one, without any further change but the introduction of more refined symmetry structures. Among these structures, we particularly mention quantum groups, the axioms of which are also naturally expressed in graph theory \cite{kauffman1995hopf}.

In future studies, generation and evolution of transient hydrodynamic knots and less-viscous liquids could help elucidate the reconnection events that have escaped previous investigations \cite{kleckner2013creation}. Synthesis of focused acoustic vortices \cite{baudoin2021orbital} or more advanced acoustic holograms \cite{riaud2015anisotropic,marzo2015holographic,melde2016holograms} could allow manufacturing complex helicity fields. Finally, combining multiple acoustic vortex transducers may enable the study of hydrodynamic knot networks as an analog to turbulence phenomena, quark-gluon plasma \cite{buniy2004glueballs} and topological quantum fields.

\section{Methods}
\subsection{Transducers preparation}
The spiraling IDTs are prepared from 1 mm thick 3'' diameter X-cut lithium niobate wafer coated by Physical Vapor Deposition with 5 nm of titanium (adhesion layer) and 100 nm of gold. The electrode pattern is then obtained by standard photolithography and wet etching. Furthermore, to ensure the droplet is accurately positioned at the center of the transducer, we pin the droplet contact line with gold ring coated with a hydrophobic surface adsorbed monolayer of 1-dodecanethiol. 

\subsection{Laser vibrometry}
The vibrations are measured with a heterodyne Mach-Zehnder vibrometer \cite{royer1985improved,royer1986optical}. To avoid reflections from the substrate edges,  we used 20-periods long bursts of a 20 MHz excitation signal. The signal generator output is set to 500 mV$_{pp}$ and the signal is then amplified by a 40 dB power amplifier (ZHL-5W-1+, minicircuits, USA) and fed to the transducer. The vibration amplitude is recorded for each individual point (see SI) and the acoustic field is then numerically reconstructed.

\subsection{General defocusing particle tracking}
In all experiments, we used a 90 w\% glycerol/water mixture (viscosity: $\mu$ = 156 mPa.s) seeded with 5 $\mu$m polystyrene fluorescent microspheres (Huge Biotechnology, Wuhan, China). In each experiment, we drop 2 $\mu$L of this mixture at the center of the transducer, and use the same electrical setup as for the vibrometry experiments, except that the excitation signal is a continuous sinusoidal signal with a 50 mV$_{pp}$ and 80 mV$_{pp}$ amplitude for the $\mathcal{W}_0$ and $\mathcal{W}_{15}$, respectively. The smaller amplitudes are chosen to ensure that the droplet surface does not deform. The reflections from the substrate edge are minimized by placing a large PDMS around the transducer \cite{raghavan2010particle}. The particles motion is recorded for 5402 frames at 5 fps ($\mathcal{W}_0$) and 20 fps ($\mathcal{W}_{15}$). After the experiment, the particle trajectories are obtained with GDPTlab and post-processed with Python. Finally, the false vectors (outside the droplet) are masked with an image editor (Inkscape). Unmasked pictures are available in SI.

\subsection{Acoustic streaming simulation}
The acoustic streaming is simulated in a similar fashion to \cite{riaud2017influence}. Briefly, we build a 2D-axisymmetric model and solve first the acoustic field and then the hydrodynamic field in a time-invariant geometry (we assume that the SAW amplitude is small enough to neglect the droplet deformation). In the acoustic model, the boundary conditions are soft sound boundaries at the water-air interface and acoustic impedance combined with the incident swirling SAW at the solid-liquid interface. In the hydrodynamic model, the acoustic streaming is modeled as an external force given by \eq{eq: Feckart}, and the boundary conditions are no-slip at the solid wall and stress-free at the liquid-air interface. A more detailed description of the simulation is available in SI.

\subsection{Data Availability}
The data that support the findings of this study are available from the corresponding author upon reasonable request.

\bibliography{References}

\begin{acknowledgements}
	
We gratefully acknowledge Shengwei Zhao for inspiring discussions, and Rune Barnkob and Massimiliano Rossi for kindly providing DefocusTracker for the GDPT analysis. This work was supported by the National Natural Science Foundation of China with Grant No. 61874033 and No. 51950410582. AM wishes to acknowledge support by the NSFC, through the grant No. 11875113, the Shanghai Municipality, through the grant No. KBH1512299, and by Fudan University, through the grant No. JJH1512105.

\end{acknowledgements}

\section{Author contributions}
AR and JZ proposed the research, SS did the experiments, AR and SS did the simulations. AM discussed with AR about some relevant topological features from the theoretical perspective of topological quantum field theory. All the authors wrote the paper together.

\section{Competing interests}
The authors declare no competing interests.

\section{Additional information}
Supplementary information is available for this paper at XXX.


\end{document}